\documentclass[twocolumn,showpacs,aps,prl,letter,amsmath,amssymb,superscriptaddress]{revtex4-1}

\usepackage{graphicx}
\usepackage{dcolumn}
\usepackage{bm}
\usepackage{hyperref}
\usepackage{subfigure}

\usepackage{color,bbm}
\usepackage[utf8x]{inputenc}
\newcommand{\beq}{\begin{equation}}
\newcommand{\eeq}{\end{equation}}
\newcommand{\bqa}{\begin{eqnarray}}
\newcommand{\eqa}{\end{eqnarray}}
\newcommand{\nn}{\nonumber}

\newcommand{\rt}[1]{\sqrt{#1}\,}

\newcommand{\smallfrac}[2]{\mbox{$\frac{#1}{#2}$}}

\newcommand{\half}{\smallfrac{1}{2}}

\newcommand{\sq}[1]{\left[ {#1} \right]}

\newcommand{\an}[1]{\left\langle{#1}\right\rangle}
\newcommand{\tr}[1]{{\rm Tr}\sq{ {#1} }}

\newcommand{\id}{\mathbbm{1}}
\newcommand{\B}{{\mathcal B}}
\newcommand{\str}{\mathcal{S}}
\newcommand{\R}{\mathcal R}

\newcommand{\p}[1]{P_{\bm{#1}}}

\definecolor{maroon}{rgb}{0.7,0,0}

\definecolor{ngreen}{rgb}{0.3,0.7,0.3}

\definecolor{golden}{rgb}{0.8,0.6,0.1}

\begin{document}

\title{Limitations on sharing Bell nonlocality between sequential pairs of observers}

\author{Shuming Cheng}
\affiliation{The Department of Control Science and Engineering, Tongji University, Shanghai 201804, China}
\affiliation{Shanghai Institute of Intelligent Science and Technology, Tongji University, Shanghai 201804, China}
\affiliation{Institute for Advanced Study, Tongji University, Shanghai, 200092, China}


\author{Lijun Liu}
\affiliation{College of Mathematics and Computer Science, Shanxi Normal University, Linfen 041000, China}

\author{ Travis J. Baker}
\affiliation{Centre for Quantum Computation and Communication Technology (Australian Research Council),
	Centre for Quantum Dynamics, Griffith University, Brisbane, QLD 4111, Australia}

\author{Michael J. W. Hall}

\affiliation{Department of Theoretical Physics, Research School of Physics, Australian National University, Canberra ACT 0200, Australia}

\date{\today}

\begin{abstract}

We give strong analytic and numerical evidence that, under mild measurement assumptions, two qubits cannot both be recycled to generate Bell nonlocality between multiple independent observers on each side. This is surprising, as under the same assumptions it is possible to recycle just one of the qubits an arbitrarily large number of times [P. J. Brown and R. Colbeck, Phys. Rev. Lett. \textbf{125}, 090401 (2020)]. We derive  corresponding `one-sided monogamy relations' that rule out two-sided recycling for a wide range of parameters, based on a general tradeoff relation between the strengths and maximum reversibilities of qubit measurements. We also show if the assumptions are relaxed to allow sufficiently biased measurement selections, then there is a narrow range of measurement strengths that allows two-sided recycling for two observers on each side, and propose an experimental test.  Our methods may be readily applied to other types of quantum correlations, such as steering and entanglement, and hence to general information protocols involving sequential measurements.

\end{abstract}


\maketitle

\paragraph{Introduction---} It is always of interest to determine ways in which physical resources can be usefully exploited. One such resource is quantum entanglement, which is critical to information tasks such as quantum teleportation~\cite{Bennett93} and secure quantum key distribution~\cite{Ekert91}. It was recently shown, in a network scenario,  that multiple pairs of independent observers can exploit the same entangled state by weakly measuring and passing along its components~\cite{Silva15}. This has generated great interest both theoretically~\cite{Mal16,Curchod17,Tavakoli18,Bera18,Sasmal18,Shenoy19,Das19,Saha19,Kumari19,Brown20,Maity20,Bowles20,Roy20} and experimentally~\cite{Schiavon17,Hu18,Choi20,Foletto20,Feng20,Foletto21}.

As a notable example, it is possible to use two entangled qubits to generate Bell nonlocal correlations between a first observer holding the first qubit and each one of an arbitrarily long sequence of independent observers that hold the second qubit in turn~\cite{Brown20}. This recycling of the second qubit allows the first observer to implement device-independent information protocols, such as secure quantum key distribution~\cite{Ekert91,Ekert14} and randomness generation~\cite{Curchod17,Pironio10,Foletto21}, with each one of the other observers.

  \begin{figure*}[!t]
 	\centering
 	
 	\includegraphics[width=0.8
 	\textwidth]{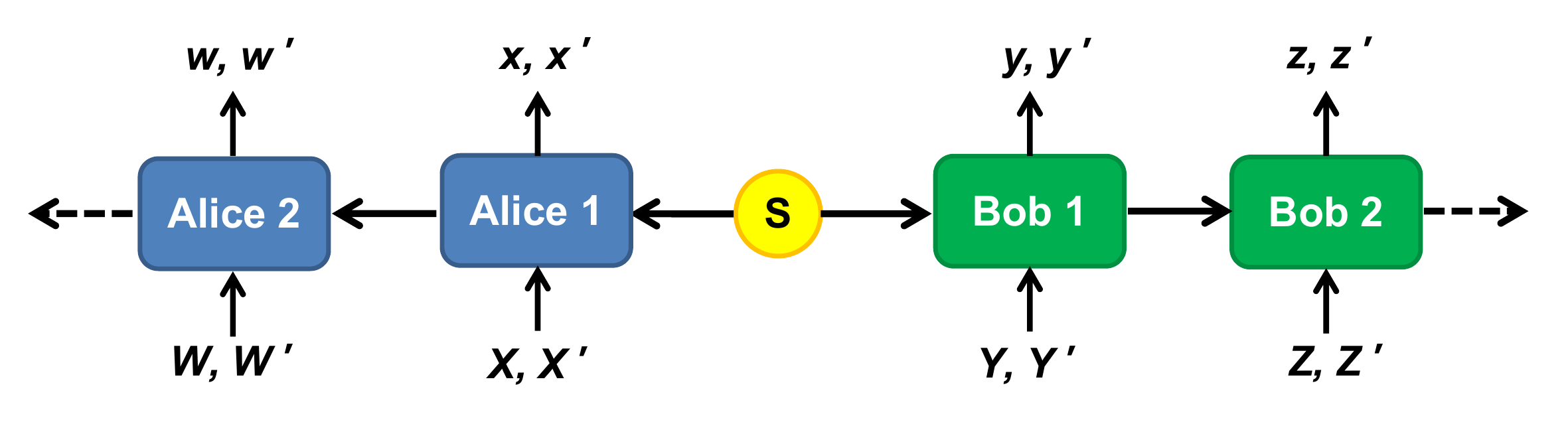}
 	
 	\caption{Sequential Bell nonlocality with multiple observers on each side. A source S generates two qubits on each run, which are received by observers Alice~1 and Bob~1 ($A_1$ and $B_1$ in the main text).  Each makes one of two local measurements on their qubit with equal probabilities (e.g., $X$ or $X'$); records their result (e.g., $x$ or $x'$); and passes their qubit onto independent observers Alice~2 and Bob~2, respectively ($A_2$ and $B_2$ in the main text). It is known that Alice~1 can demonstrate Bell nonlocality with each of an arbitrary number of Bobs in this way, via recycling of the second qubit~\cite{Brown20}. However, strong analytic and numerical evidence supports the conjecture that, surprisingly, Alice~1 and Bob~1 can demonstrate Bell nonlocality via their measurements only if Alice~2 and Bob~2 cannot, and vice versa. A similar conjecture and evidence applies to the pairs (Alice~1, Bob~2) and (Alice~2, Bob~1). Thus, the qubits cannot {\it both} be recycled to sequentially generate Bell nonlocality in this way. However, by significantly relaxing the equal-probability assumption, this limitation can be overcome to allow all four pairs to demonstrate a small degree of Bell nonlocality and thus to implement device-independent protocols such as randomness generation. 
 	}
 	\label{fig:fig1}
 \end{figure*}

In contrast, we show here that there are surprisingly strong limitations on recycling {\it both} qubits in this way, so as to generate Bell nonlocality for {\it multiple} observers on each side. This is so even for just two observers on each side (see Fig.~\ref{fig:fig1}). In particular, we give strong numerical evidence for the conjecture that if two observers independently make one of two equally-likely two-valued measurements on their qubits, as in~\cite{Brown20}, then a second pair of observers cannot observe Bell nonlocality if the first pair does. This limitation, to one-sided recycling, may be regarded as a type of sharing monogamy, which we demonstrate analytically for a wide range of parameters via corresponding `one-sided monogamy relations'.

The physical intuition behind such limitations is that if the measurements made by the first pair of observers in this scenario are sufficiently strong to demonstrate Bell nonlocality, then they are also sufficiently irreversible to leave the qubits in a Bell-local state. Correspondingly, the analytic results rely on a tradeoff between the strength and maximum reversibility of general two-valued qubit measurements, as shown below, of some interest in its own right.

We show that the validity of the conjecture only requires explicit consideration of the 16-parameter set of observables measured by the first pair of observers, on a 1-parameter class of pure initial states, making a numerical test feasible. Further,  we obtain analytic monogamy relations for two 14-parameter subsets, corresponding to the observables having either equal  strengths or orthogonal measurement directions for each side. These monogamy relations hold for all initial states with maximally-mixed marginals, and for arbitrary initial states if the observables are unbiased.

Finally, by allowing the first pair of observers to select one of their measurements with high probability ($>90\%$), we show it becomes possible for each of the observers on one side to generate Bell nonlocality with each of those on the other side, via a judicious choice of measurement strengths, and a corresponding experimental test is proposed. However, while this restores a degree of symmetry to qubit recycling, it is at the cost of a significant asymmetry in the selection of measurements, that strongly limits, e.g., the randomness that can be generated in device-independent protocols. A number of details and generalisations are left to the Supplemental Material~\cite{SM} and a forthcoming companion paper~\cite{Cheng21}.

\begin{figure*}[!t]
	\centering
	\includegraphics[width=0.9\textwidth]{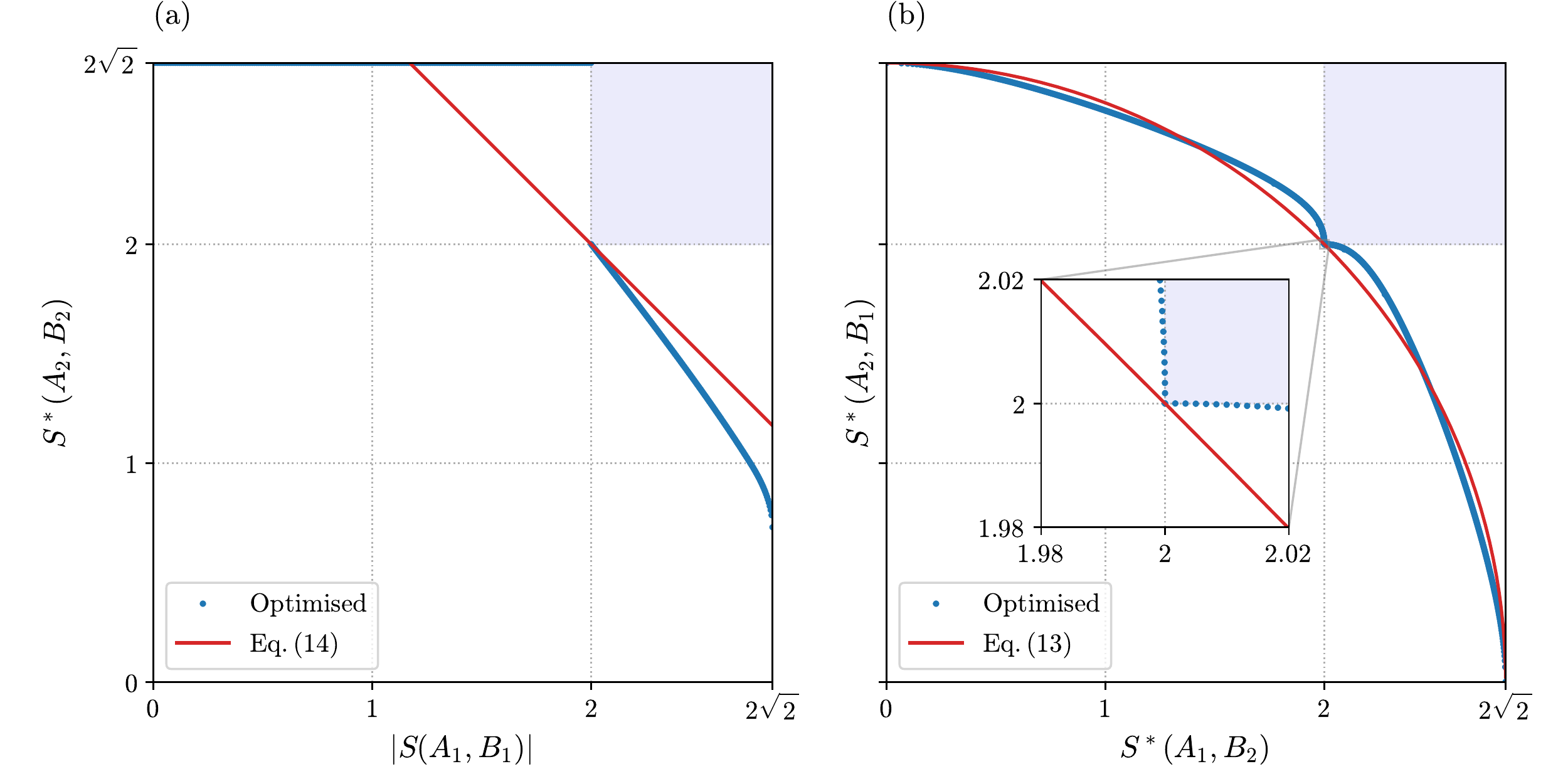}
	\caption{ One-sided monogamy for CHSH Bell nonlocality.
	The possibility of violating the CHSH inequality (a) by both $(A_1,B_1)$ and $(A_2,B_2)$, and (b) by both  $(A_1,B_2)$ and $(A_2,B_1)$, is tested via the values of $|S(A_1,B_1)|$  and the proxy quantities $S^*(A_1, B_2), S^*(A_2, B_1), S^*(A_2,B_2)$ in Eqs.~(\ref{chsh}), (\ref{smax22})--(\ref{s21}).
	In both cases, blue data points correspond to a numerically maximised value of the quantity on the $y$-axis under the constraint that the quantity on the $x$-axis achieves a minimum given value, which is varied over $[0,2\sqrt{2}]$.
The absence of any values falling in the shaded regions supports the one-sided monogamy conjecture for qubit recycling (see main text).
For unbiased observables with equal strengths $\str_X=\str_{X'}$ and $\str_Y=\str_{Y'}$ the one-sided monogamy relations in Eqs.~(\ref{monog1}) and~(\ref{monog2}) are analytically satisfied, corresponding to values below the red lines. 
Similar relations hold for orthogonal spin directions and/or zero Bloch vectors (see main text). 
}
	\label{fig:fig2}
\end{figure*}

\paragraph{One-sided monogamy conjecture---}
Before proceeding to details, we formally state the main conjecture and preview the numerical evidence. First, if an observer $A$ ($B$) measures either of two observables $X$ or $X'$ ($Y$ or $Y'$), with outcomes labelled by $\pm1$, then Bell nonlocality is characterised by the value of the Clauser-Horne-Shimony-Holt (CHSH) parameter~\cite{Clauser69,Brunner14}
\beq \label{chsh}
S(A,B):= \langle XY\rangle+\langle XY'\rangle+\langle X'Y\rangle - \langle X'Y'\rangle .
\eeq
In particular, a violation of the CHSH inequality $S(A,B)\leq2$ implies there is no local hidden variable model for the correlations between the measurement outcomes, guaranteeing the security of device independent protocols such as quantum key distribution and randomness generation. Our results strongly support the following conjecture, that significantly limits the recycling of qubits used for the generation of Bell nonlocality.\\
{\bf Conjecture:} {\it If observers $A_1$ and $B_1$ independently make one of two equally-likely measurements on a first and second qubit, respectively, and the qubits are passed on to observers $A_2$ and $B_2$, respectively, then the pairs $(A_j,B_k)$ and $(A_{j'},B_{k'})$ can each violate the CHSH inequality only if they share a common observer, i.e., 	
}
\beq \label{conjecture}
S(A_j,B_k),\, S(A_{j'},B_{k'})>2~~{\rm  only~if}~ j=j'~{\rm or}~k=k'.
\eeq
Thus, the conjecture asserts that sequential Bell nonlocality is only possible via a fixed observer on one side in this scenario, as in~\cite{Brown20}, but {\it not} for multiple observers on each side. In particular, it implies that at most one of the pairs $(A_1,B_1)$ and $(A_2,B_2)$ can violate the CHSH inequality, and similarly at most one of the pairs $(A_1,B_2)$ and $(A_2,B_1)$.  Here  $A_1$ corresponds to Alice~1 in Fig.~\ref{fig:fig1}, etc. Numerical evidence for the conjecture is illustrated in Fig.~\ref{fig:fig2}. We note the conjecture does not generally extend to, e.g., Bell inequalities with more measurements per observer on higher-dimensional systems ~\cite{adan}.

\paragraph{Qubit measurements---}
To proceed, consider measurement of a (generalised) two-valued observable described by a positive operator valued measure (POVM) $\{X_+,X_-\}$. Thus, $X_\pm\geq 0$ with $X_++X_-=\id$, and the observable is equivalently represented by the operator $X:= X_+ - X_-$, with $-\id\leq X\leq \id$. For qubits, $X$ can be decomposed as
\beq
X =  \B \id + \str\bm \sigma\cdot\bm x , \label{measurement}
\eeq
with respect to the Pauli spin operator basis $\boldsymbol{\sigma}\equiv (\sigma_1, \sigma_2, \sigma_3)$, with $\str\geq 0$ and $|\bm x|:= (\bm x\cdot \bm x)^{1/2} =1$. Here $\B$ is the {\it outcome bias} of the observable, $\str$ is its {\it strength}~\cite{Shenoy19} (or information gain~\cite{Silva15}), and $\bm x$ is a direction associated with the observable. For projective observables one has $\B=0$ and $\str=1$, while for the trivial observable with POVM $\{\id,0\}$ one has $\B=1$ and $\str=0$. More generally, $\B$ is the difference of the $+1$ and $-1$ outcome probabilities for the maximally-mixed state $\rho=\half\id$, and $|\B|+ \str\leq 1$.  

In the context of sequential measurements, as in Fig.~\ref{fig:fig1}, the effect of a measurement on the subsequent state of a qubit is important. For example, an observation of $X\equiv\{X_+,X_-\}$ can  be implemented by the square-root measurement that takes 
the state $\rho$ to 
\begin{align}
	\phi(\rho):=X_+^{1/2}\rho X_+^{1/2} + X_-^{1/2}\rho X_-^{1/2}.
\end{align}
More generally, any measurement of  $X$ takes $\rho$ to $\phi_G(\rho) = \phi_+(X_+^{1/2}\rho X_+^{1/2}) + \phi_-(X_-^{1/2}\rho X_-^{1/2})$, for two quantum channels $\phi_+$ and $\phi_-$~\cite{Brown20}, equivalent to first carrying out the square-root measurement and then applying a quantum channel that may depend on the outcome. Now, a quantum channel is reversible if and only if it is unitary, implying the square-root measurement map $\phi$ can be recovered from $\phi_G$ only if $\phi_\pm$ are unitary transformations, and indeed the same unitary transformation if the outcome is unknown (as is the case for independent sequential measurements). Hence, the square-root measurement $\phi$ is optimal, in the sense of being the maximally reversible measurement of $X$ (up to a unitary transformation), and we follow~\cite{Brown20} in confining attention to such measurements.  

To quantify the degree of maximum reversibility, we note that explicit calculation gives~\cite{SM}
\begin{align}
	\phi(\rho)
	&= \p{x}\rho\p{x} + \p{-x}\rho\p{-x}+ \R\left(\p{x}\rho\p{-x}+\p{-x}\rho\p{x}\right),
	\label{phirho}
\end{align}
where $P_{\bm x}=\half(\id+\bm\sigma\cdot\bm x)$ denotes the projection onto unit spin direction $\bm x$, and 
\beq 
\R := \half\sqrt{(1+\B)^2-\str^2} + \half\sqrt{(1-\B)^2-\str^2} . \label{rdef}
\eeq
Thus, the off-diagonal elements of $\rho$ in the $\bm \sigma\cdot \bm x$ basis are scaled by $\R$, with $\R=0$ for projective observables ($\B=0, \str=1)$, and $\R=1$ for trivial observables ($\str=0$). Hence, $\R$ is a natural measure of the {\it maximum reversibility} associated with the measurement of a given observable (it also upper bounds the `quality factor' $\mathcal F$ of a class of unbiased weak qubit measurements~\cite{Silva15,SM}). For convenience we will often simply refer to $\R$ as the reversibility in what follows.

\paragraph{Tradeoff between strength and reversibility---} 
Equation~(\ref{rdef}) is a general relation connecting outcome bias, strength, and maximum reversibility, and implies the fundamental tradeoff relation~\cite{SM}
\beq
\R^2+\str^2 \leq 1 ,\label{tradeoff}
\eeq
between reversibility and strength. Equality holds for any unbiased observable, i.e., for $\B=0$. This tradeoff is very useful for studying the shareability of Bell nonlocality via sequential measurements, and may be used to reinterpret  the information-disturbance relation given in~\cite{Banaszek01} (in the case of qubit measurements)~\cite{SM}. Equations~(\ref{phirho}) and~(\ref{tradeoff}) also suggest a natural definition of ``minimal decoherence'', ${\cal D}=\sqrt{ 1 - \R^2}\geq \str$.

\paragraph{Simplifying the conjecture---} 
Fortunately, the validity of the conjecture does not require explicit consideration of all possible initial states, nor of all possible observables measured by the four observers. For example, since the CHSH parameters $S(A_j,B_k)$ in Eq.~(\ref{chsh}) are convex-linear in the initial state $\rho$ shared by $A_1$ and $B_1$, only pure initial states $\rho=|\psi\rangle\langle\psi|$ need be considered to test the joint ranges of the parameters for any given measurements. We can further restrict to the 1-parameter class
\beq \label{pure}
|\psi\rangle= \cos\alpha |0\rangle\otimes |0\rangle + \sin \alpha |1\rangle\otimes |1\rangle,~~~~ \alpha\in[0,\pi/2]
\eeq
in the $\sigma_3$-basis $\{|0\rangle,|1\rangle\}$, as Bell nonlocality is invariant under local unitaries~\cite{Brunner14}. 

Moreover, once $A_1$  and $B_1$'s observables $X,X',Y,Y'$ have been specified, the optimal choices for $A_2$ and $B_2$ are uniquely determined, on each run, by which observer on the other side they are trying to generate Bell nonlocality with. 
For example, it follows from  Eq.~(\ref{phirho}) that if $A_1$ and $B_1$ choose between their measurements with equal probabilities, as per the conjecture, and $T$ denotes the initial spin correlation matrix (with coefficients $T_{jk}:=\tr{\rho\,\sigma_j\otimes\sigma_k}$), then the  correlation matrix shared by $A_2$ and $B_2$ is $KTL$, with~\cite{SM}
\beq \label{barka}
K:= \half(\R_{X}+\R_{X'})I_3 + \half(1-\R_{X})\bm x\bm x^\top + \half(1-\R_{X'})\bm x'\bm x'^\top 
\eeq
\beq \label{barkb}
L:= \half(\R_{Y}+\R_{Y'})I_3 + \half(1-\R_{Y})\bm y\bm y^\top + \half(1-\R_{Y'})\bm y'\bm y'^\top \nn
\eeq 
(here $\R_X$ denotes the reversibility associated with observable $X$, etc.).  It follows immediately from the Horodecki criterion that $A_2$ and $B_2$ can violate the CHSH inequality if and only if~\cite{Horodecki95}
\beq \label{smax22}
S^*(A_2,B_2):= 2\sqrt{s_1(KTL)^2 + s_2(KTL)^2} > 2,
\eeq
where $s_1(M)$ and $s_2(M)$ denote the two largest singular values of matrix $M$. 

Similarly, it can be shown that $(A_1,B_2)$ and $(A_2,B_1)$ can violate the CHSH inequality if and only if~\cite{SM}
\begin{align} 
	S^*(A_1, B_2) &:= \left|  (\B_X+\B_{X'})L\bm b + LT^\top(\tilde{\bm x}+\tilde{\bm x}')\right| \nn\\
	&~~~+\left| (\B_X-\B_{X'})L\bm b + LT^\top(\tilde{\bm x}-\tilde{\bm x}')\right| \label{s12} \\
	S^*(A_2, B_1) &:= \left|  (\B_Y+\B_{Y'})K\bm a + KT(\tilde{\bm y}+\tilde{\bm y}')\right| \nn\\
	&~~~+\left| (\B_Y-\B_{Y'})K\bm a + KT(\tilde{\bm y}-\tilde{\bm y}')\right| \label{s21}
\end{align}
are greater than 2, respectively, where $\tilde{\bm x}:=\str_X\bm x, \tilde{\bm x}':=\str_{X'}{\bm x}'$, etc, and $\bm a$ and $\bm b$ are the initial Bloch vectors of the first and second qubits. 

Hence, to verify the conjecture one need only consider the values of $S(A_1,B_1)$ and the proxy quantities $S^*(A_1, B_2), S^*(A_2, B_1), S^*(A_2,B_2)$, over the 16-parameter set of observables $X,X',Y,Y'$ and a 1-parameter set of initial states. Moreover, if the observables are unbiased (i.e., $\B_X=\B_{X'}=\B_Y=\B_{Y'}=0$), 
then it may be shown that the conjecture need only be verified for a 9-parameter set of observables on a single maximally entangled state~\cite{SM}. Note that prior work~\cite{Silva15,Mal16,Curchod17,Tavakoli18,Bera18,Sasmal18,Shenoy19,Das19,Saha19,Kumari19,Brown20,Maity20,Bowles20, Roy20,Schiavon17,Hu18,Choi20,Foletto20,Feng20,Foletto21} is restricted to unbiased observables.

\paragraph{Evidence for conjecture---}Our numerical evidence is based on performing optimisations of the proxy quantities $S^*(A_2,B_2)$ and $S^*(A_2, B_1)$ under the constraint that the respective quantities $S(A_1,B_1)$ and $S^*(A_1, B_2)$ achieve a minimum fixed value. These maximisations over the 17-dimensional parameter space were computed using a constrained differential evolution algorithm~\cite{Sto97,Lam02} implemented in the \textsc{SciPy} library~\cite{Vir20} 
(see Section IV of the Supplemental Material~\cite{SM} for details). All data points found through optimisation---illustrated as the blue dots in Fig.~\ref{fig:fig2}---lie strictly outside the grey regions, providing strong evidence the conjecture holds.

Moreover, we can directly prove the conjecture in a number of scenarios, using tradeoff relation~(\ref{tradeoff}). For example, for the case of equal  strengths for each side, i.e., $\str_X=\str_{X'}$ and $\str_{Y}=\str_{Y'}$, the one-sided monogamy relation
\beq \label{monog1}
S^*(A_1, B_2)^2 + S^*(A_2,B_1)^2 \leq 8
\eeq
holds for all states $\rho$ with maximally mixed marginals, i.e., $\bm a=\bm b=0$,  and for arbitrary states if the observables are unbiased~\cite{SM}, implying Eq.~(\ref{conjecture}) holds for these pairs.
The upper bound corresponds to the red curve on the right hand panel of Fig.~\ref{fig:fig2}. This relation also holds for the alternative case of orthogonal measurement directions on each side, i.e., $\bm x\cdot\bm x'=\bm y\cdot\bm y'=0$~\cite{SM}.  

One-sided monogamy relations for the pairs $(A_1,B_1)$ and $(A_2,B_2)$ can also be derived, such as
\beq \label{monog2}
|S(A_1, B_1)| + S^*(A_2,B_2) \leq 4
\eeq
for the case of unbiased observables and equal strengths. The upper bound corresponding to the red curve on the left hand panel of Fig.~\ref{fig:fig2}, and can be improved to $16/(3\sqrt{2})$ for the case of orthogonal directions. However, since these relations require considerably more work (including a significant generalisation of the Horodecki criterion), they are only derived under {\it joint} strength  and orthogonality assumptions in~\cite{SM}, with the general cases left to a companion paper~\cite{Cheng21}.

\paragraph{Sequential Bell nonlocality via biased measurement selections---} 
The conjecture and above results require that observers $A_1$ and $B_1$ each select their measurements with equal probabilities. However, if they instead select between making a relatively weak measurement, with sufficiently high probability, and a relatively strong measurement, with correspondingly low probability, then the average disturbance to their state  can be small enough to allow {\it all four pairs} to generate Bell nonlocality.

To show this, suppose that $A_1$ measures $X$ and $X'$ with probabilities $1-\epsilon$ and $\epsilon$, and $B_1$ similarly measures $Y$ and $Y'$ with probabilities $1-\epsilon$ and $\epsilon$, on a singlet state (i.e., $T=-I$). Suppose further  that $X$ and $Y$ are measured with strengths $\str_X=\str_Y=\str$ and reversibilities $\R_X=\R_Y=\R$;  that $X'$ and $Y'$ are projective, with strengths $\str_{X'}=\str_{Y'}=1$ and reversibilities $\R_{X'}=\R_{Y'}=0$; and the measurement directions are the optimal CHSH directions~\cite{Clauser69} (i.e., $\bm x$ and $\bm x'$ are orthogonal with $\bm y=(\bm x+\bm x')/\sqrt{2}$ and $\bm y'=(\bm x-\bm x')/\sqrt{2}$).  Then 
$|S(A_1,B_1)|= (\str+1)^2/\sqrt{2}$ from Eq.~(\ref{chsh}),
implying $A_1$ and $B_1$ can violate the CHSH inequality only if this is larger than 2, i.e., only if
\beq \label{rplus}
\str > 8^{1/4}-1 \sim 0.682.
\eeq
Further, the state shared by $A_2$ and $B_2$ has spin correlation matrix $T=-K_\epsilon L_\epsilon$, with
$K_\epsilon = (1-\epsilon)[ \R I  + (1-\R) \bm a\bm a^\top ]   
+ \epsilon  \bm a'\bm a'^\top$,
and a similar expression for $L_\epsilon$ with $\bm x,\bm x'$ replaced by $\bm y,\bm y'$, implying via Eq.~(\ref{smax22}) that $A_2$ and $B_2$ can violate the CHSH inequality only if~\cite{SM}
\beq \label{rmin}
\R > \frac{[\sqrt{2}-(1-\epsilon)^2]^{1/2}-\epsilon}{1-\epsilon}> (\sqrt{2}-1)^{1/2} \sim 0.644  .
\eeq
Equations~(\ref{tradeoff}), (\ref{rplus}) and (\ref{rmin}) yield a narrow range of strengths, $0.682<\str<0.765$, over which both $(A_1,B_1)$ and $(A_2,B_2)$ can generate Bell nonlocality by recycling both qubits. Further, this range constrains the  probability of making the projective measurement to~\cite{SM}
\beq
\epsilon<\epsilon_{\max}\sim 7.9\% .
\eeq
The pairs $(A_1,B_2)$ and $(A_2,B_1)$ can also generate Bell nonlocality under these conditions~\cite{SM}.

\paragraph{Discussion---} 
Strong numerical and analytic evidence has been given to support an unexpected one-sided monogamy conjecture, that limits sequential violation of the CHSH inequality to one-sided qubit recycling if observers make  unbiased measurement selections. Conversely, allowing sufficiently biased selections permits a narrow range of measurement strengths within which two-sided qubit recycling is possible. We propose testing the latter experimentally, as it in principle permits four independent pairs of observers to generate Bell nonlocality, and hence to carry out device independent quantum information protocols such as randomness generation, via recycling of a two-qubit state. Generalisations of our methods are given in~\cite{Cheng21}, and we expect these methods can also be readily applied to the sequential sharing of quantum properties such as entanglement~\cite{Horodecki09}, Einstein-Podolsky-Rosen steering~\cite{Uola20}, and random access codes~\cite{Mohan19,Anwer20,Das21}. We also hope to find a more rigorous justification for restricting to square root measurements.

\acknowledgements We thank Yong Wang, Xinhui Li, Jie Zhu, Mengjun Hu, Ad\'an Cabello and two anonymous referees for helpful discussions and comments. S. C.~is supported by the Fundamental Research Funds for the Central Universities ( No.~22120210092) and the National Natural Science Foundation of China (No.~62088101). L. L.~is supported by National Natural Science Foundation of China (No.~61703254). T. J. B. is supported by the Australian Research Council Centre of Excellence CE170100012, and acknowledges the support of the Griffith University eResearch Service \& Specialised Platforms Team and the use of the High Performance Computing Cluster ``Gowonda'' to complete this research.  \\
{\it Note:} Monogamy for the pairs $(A_1,B_1), (A_2,B_2)$, for the very special case of  measurements of unbiased observables on a singlet state in the optimal CHSH directions and with equal strengths for each side, has been noted independently in a recent work~\cite{Jie21}.

~~
\newpage

\setcounter{equation}{0}
\renewcommand{\theequation}{S.\arabic{equation}}

\setcounter{figure}{0}
\renewcommand{\thefigure}{S\arabic{figure}}

\setcounter{page}{1}
\renewcommand{\thepage}{Supplemental Material -- \arabic{page}/10}

\section{\large{SUPPLEMENTAL MATERIAL}}

\section{I. Bias, strength and maximum reversibility of two-valued qubit measurements}

\subsection{A. Measurement strength and outcome bias}

Recall from the main text that a general two-valued qubit observable, with POVM $\{X_+,X_-\}$, is equivalently represented by the operator
\beq \label{xrep}
X = X_+-X_- = {\cal B} \id + {\cal S}\bm \sigma\cdot\bm x ,
\eeq
where $\str\geq 0$ and $\B$
denote the {\it strength}  and {\it outcome bias} of the observable, respectively, and $\bm x$ is the associated measurement direction, with unit norm  $|\bm x|:=  (\bm x\cdot\bm x)^{1/2}  =1$. The POVM elements are determined uniquely by $X$ via $X_\pm=\half(\id\pm X)$, and the positivity requirement $X_\pm\geq 0$ is equivalent to $-\id\leq X\leq \id$, i.e., to the condition 
\beq 
|\B|+\str\leq 1.
\eeq
on the strength and bias. We note that $\str$ is also referred to as  `information gain' and denoted by $G$~[3]. However, we prefer to use the alternative term `strength'~[9], in part to distinguish it from the average information gain $G$ defined by Banaszek~[30] (see also Sec.~1.C below). Another possible terminology is `sharpness'~[31].  

It is convenient for later purposes to write $X$ in the form
\beq \label{eig}
X := x_+P_{\bm x} + x_-P_{-\bm x},
\eeq
where 
\beq
P_{\bm x}:=\frac{\id+\bm\sigma\cdot\bm x}{2} 
\eeq
denotes the projection onto spin direction $\bm x$. Hence,
\beq \label{bias} 
x_\pm=\B\pm\str, 
\eeq
and
\beq \label{Xpm}
X_\pm = \frac{1\pm x_+}{2}\p{x}+\frac{1\pm x_-}{2}\p{-x} .
\eeq

\subsection{B. Maximum reversibility}

To obtain Eqs.~(5) and~(6) of the main text, note from Eq.~(\ref{Xpm}) that
\beq \label{root}
X_\pm^{1/2} = \sqrt{\frac{1\pm x_+}{2}}\p{x}+\sqrt{\frac{1\pm x_-}{2}}\p{-x}.
\eeq
Hence, the square-root measurement operation corresponding to $X$ takes state $\rho$ to the state
\begin{align}
\phi(\rho)&:=X_+^{1/2}\rho X_+^{1/2} + X_-^{1/2}\rho X_-^{1/2}\nn\\
	&= \left(\frac{1+x_+}{2}+\frac{1-x_+}{2}\right)\p{x}\rho\p{x}\nn\\
	&~+ \left(\frac{1+x_-}{2}+\frac{1-x_-}{2}\right)\p{-x}\rho\p{-x}\nn\\
	&~+\half\sqrt{(1+x_+)(1+x_-)}\left(\p{x}\rho\p{-x}+\p{-x}\rho\p{x}\right)\nn\\
	&~+\half\sqrt{(1-x_+)(1-x_-)}\left(\p{x}\rho\p{-x}+\p{-x}\rho\p{x}\right)\nn\\
	&= \p{x}\rho\p{x} + \p{-x}\rho\p{-x}+ \R\left(\p{x}\rho\p{-x}+\p{-x}\rho\p{x}\right),
	\label{rhoprime}
\end{align}
as per Eq.~(5), with the maximum reversibility $\R$ given by
\beq \label{reverse}
\R :=\half\sqrt{(1+x_+)(1+x_-)} + \half\sqrt{(1-x_+)(1-x_-)} .
\eeq
Equation~(\ref{bias}) then yields 
\beq 
\R = \half\sqrt{(1+\B)^2-\str^2} + \half\sqrt{(1-\B)^2-\str^2}  \label{reversibility}
\eeq
as per Eq.~(6) of the main text.

The interpretation of $\R$ as the {\it maximum} reversibility of any measurement of $X$ is directly supported by a class of unbiased weak qubit measurements considered by Silva {\it et al.}. For these measurements, comparing Eq.~(\ref{xrep}) above with Eqs.~(3) and~(42) of~[3], 
\beq
\B=0,\qquad \str = \frac{1-e^{-2{\rm Re}(a)}}{1+e^{2{\rm Re}(a)}},
\eeq
where $a$ is a complex parameter with nonnegative real part. This corresponds to a maximum reversibility 
\beq
\R = \sqrt{1-\str^2} = \frac{2e^{-{\rm Re}(a)}}{1+e^{2{\rm Re}(a)}} 
\eeq
via Eq.~(\ref{reversibility}).
Further, from Eqs.~(4) and~(44) of~[3] the postmeasurement state is given by
\beq
\phi_a(\rho) := \p{x}\rho\p{x} + \p{-x}\rho\p{-x}+ {\mathcal F}\left(\p{x}\rho\p{-x}+\p{x}\rho\p{-x}\right)
\eeq
where  $\mathcal F$ is  the `quality factor' defined by
\beq
{\mathcal F}:= \frac{2e^{-{\rm Re}(a)} |\cos {\rm Im}(a)|}{1+e^{2{\rm Re}(a)}}  .
\eeq
Thus, the off-diagonal elements of the post-measurement state are scaled by a factor $\mathcal F\leq\R$ (with equality for ${\rm Im}(a)=0$), and so are clearly less reversible than the square-root measurement in general, as expected. It further follows that optimal measurements of this type, with ${\mathcal F}={\mathcal F}_{\max}=\R$, correspond to square root measurements.

\subsection{C. Fundamental tradeoff relation}
\label{sec:tradeoff}

Squaring each side of Eq.~(\ref{reversibility}), then rearranging and squaring again, leads to the tradeoff relation
\beq \label{rssum}
\R^2+\str^2 = 1- \B^2(\frac{1}{\R^2}-1)\leq 1 
\eeq
between strength and reversibility, as per Eq.~(7) of the main text.
This tradeoff is not only very helpful for studying the shareability of Bell nonlocality via sequential measurements, but is also of interest more generally.

For example, for a general quantum measurement on a $d$-dimensional  system, Banaszek defines a corresponding `mean operation fidelity' $F$, related to the disturbance caused by the measurement, and a `mean estimation fidelity' $G$, related to the average information gain or quality of the measurement, and shows that these satisfy the general information-disturbance relation~[29]
\beq
\sqrt{F-\frac{1}{d+1}}\leq\sqrt{G-\frac{1}{d+1}}+\sqrt{(d-1)(\frac{2}{d+1}-G)}. \nn
\eeq
For a two-valued qubit measurement with Kraus operators $M_\pm$, corresponding to the POVM $\{X_\pm=M^\dagger_\pm M_\pm\}$, these quantities may be calculated explicitly, yielding
\begin{align} 
6F &= 2 + \left|\tr{M_+}\right|^2 + \left|\tr{M_-}\right|^2 \nn\\
&= 2 + \left|\tr{U_+X_+^{1/2}}\right|^2 + \left|\tr{U_-X_-^{1/2}}\right|^2 \nn\\
&\leq 2 + \tr{X_+^{1/2}}^2 + \tr{X_-^{1/2}}^2 \nn\\
&= 2\R + 4 \label{fdef}
\end{align}
(where the second line uses the polar decomposition $M_\pm=U_\pm X_\pm^{1/2}$ for unitary operators $U_\pm$ and the last line follows via Eqs.~(\ref{root}) and~(\ref{reverse})), and
\beq
6G = 2 + \lambda_{\max}(X_+) + \lambda_{\max}(X_-) = \str+3,
\eeq
where $\lambda_{\max}(M)$ denotes the maximum eigenvalue of $M$. Thus, $F$ and $G$ can be reinterpreted in terms of  the maximum reversibility and strength of the measurement for this case (note also the {\it maximum} reversibility property of $\R$ is emphasised by the inequality in Eq.~(\ref{fdef}), which is saturated for the square-root measurement $M_\pm=X_\pm^{1/2}$). Further, our tradeoff relation~(\ref{rssum}) for $\R$ and $\str$ can be rewritten as
\beq \nn
2+2\R \leq 2+2\sqrt{1-\str^2} = \left(\sqrt{1+\str}+\sqrt{1-\str}\right)^2,
\eeq
which, on taking the square root and substituting the above expressions for $F$ and $G$, yields
\beq
\sqrt{6F-2} \leq \sqrt{6G-2} + \sqrt{4-6G}.
\eeq
Thus, the fundamental tradeoff relation is equivalent to Banaszek's information-disturbance relation for qubit measurements. 

Generalisations and further applications of tradeoff relation~(\ref{rssum}) will be discussed elsewhere~[26].

\section{II. Spin correlation matrix, Bloch vectors, and the forms of $K$ and $L$}

The spin correlation matrix $T$ and Bloch vectors $\bm a, \bm b$ of a two-qubit state $\rho$ are defined by
\beq
T_{jk}:=\tr{\rho (\sigma_j\otimes\sigma_k)},
\eeq
\beq
a_j:=\tr{\rho(\sigma_j\otimes\id)}, ~~b_j:=\tr{\rho(\id\otimes\sigma_j)},
\eeq
for $j,k=1,2,3$.  To calculate the effect of a maximally reversible measurement of $X$ by $A_1$ on these quantities,  note first that substituting $\p{x}=\half(\id+\bm\sigma\cdot\bm x)$ in Eq.~(\ref{rhoprime}) gives
\begin{align}
	\phi(\rho)= \half(\rho+\bm\sigma\cdot\bm x\rho\bm\sigma\cdot\bm x) +\half\R(\rho - \bm\sigma\cdot\bm x\rho\bm\sigma\cdot\bm x).
\end{align}
Thus, $\phi(\id)=\id$ (i.e., the map is unital), and using the identity
\begin{align}
	(\bm\sigma\cdot\bm x)\sigma_j(\bm\sigma\cdot\bm x)&= x_kx_l\sigma_k\sigma_j\sigma_l\nn \\
	&=x_kx_l\sigma_k(\delta_{jl} + i\epsilon_{jlm}\sigma_m) \nn\\
	&=x_j(\bm \sigma\cdot\bm x)+ ix_kx_l \epsilon_{jlm}(\delta_{km}+i\epsilon_{kmn}\sigma_n)\nn\\
	&=x_j(\bm \sigma\cdot\bm x) +x_kx_l \epsilon_{jlm}\epsilon_{knm}\sigma_n\nn\\
	&=x_j(\bm \sigma\cdot\bm x) +x_kx_l(\delta_{jk}\delta_{ln}-\delta_{jn}\delta_{kl})\sigma_n\nn\\
	&=2x_j(\bm \sigma\cdot\bm x) - \sigma_j ,
\end{align}	
with summation over repeated indices, gives 
\beq
\phi(\sigma_j)  = x_j(\bm \sigma\cdot\bm x) + \R(\sigma_j -x_j\bm \sigma\cdot\bm x).
\eeq
Thus, noting from Eq.~(\ref{rhoprime}) that the map is self-dual, i.e., $\tr{\phi(M)N}=\tr{M\phi(N)}$, a measurement of $X$ on the first qubit of a two-qubit state $\rho$ changes the spin correlation matrix $T$ to $T^X$, with 
\begin{align}
	T^X_{jk} &= \tr{ (\phi\otimes I)(\rho)\,(\sigma_j\otimes\sigma_k)}
		=\tr{\rho \,\phi(\sigma_j)\otimes \sigma_k}\nn\\
	&= \langle [  x_j(\bm \sigma\cdot\bm x) + \R(\sigma_j -x_j\bm \sigma\cdot\bm x)]\otimes\sigma_k\rangle\nn\\
	&= \R\langle \sigma_j\otimes\sigma_k\rangle +(1-\R)x_jx_l\langle \sigma_l\otimes\sigma_k\rangle \nn\\
	&= \R T_{jk} + (1-\R)x_jx_lT_{lk}. 
\end{align}
Hence, 
\beq \label{tx}
T^X = K^XT,\qquad K^X:= \R_{X} I_3 +(1-\R_{X})\bm x\bm x^\top ,
\eeq
where $I_3$ is the $3\times3$ identity matrix, and we have replaced $\R$ by $\R_{X}$ to indicate we are referring to the measurement $X$. One similarly finds that the Bloch vector $\bm a$ changes to $T^X\bm a$, while the Bloch vector $\bm b$ of the second qubit is, of course, unchanged by a measurement on the first. 

Similarly, if observer $B_1$ measures the POVM $Y\equiv\{Y_+,Y_-\}$, then the spin correlation matrix changes to
\beq \label{ty}
T^Y = TL^Y,\qquad L^Y:= \R_{Y} I_3 +(1-\R_{Y})\bm y\bm y^\top,
\eeq
and the Bloch vector of the second qubit changes to $L^Y\bm b$. Further, if {\it both} observers make a measurement, then the spin matrix and Bloch vectors $\bm a,\bm b$  transform to
\beq \label{txy}
T^{XY} = K^XTL^Y, ~~  K^X\bm a,~~ L^Y\bm b,
\eeq
respectively.

Finally, if $A_1$ and $B_1$ each measure one of two POVMs, $X$ or $X'$ and $Y$ or $Y'$, with equal probabilities, it follows that $K^X$ and $L^Y$ above are replaced by the `average' matrices $K$ and $L$ defined by
\begin{align} 
K&:= \half(K^X+K^{X'}),~~~ L&:= \half(L^Y+L^{Y'}),
\label{barka}
\end{align}
as per Eq.~(9) of the main text.

\section{III. Simplifying the conjecture}

\subsection{A. Derivation of $S^*(A_1,B_2)$ and $S^*(A_2,B_1)$}

We first demonstrate that, for given observables $X,X',Y,Y'$ measured by $A_1$ and $B_1$, that $A_2$ and $B_1$ can violate the CHSH inequality if and only if $S^*(A_2,B_1)>2$ as per Eq.~(12) of the main text,  and similarly that $A_1$ and $B_2$ can violate the CHSH inequality if and only if $S^*(A_1,B_2)>2$. This greatly simplifies obtaining numerical and analytic evidence for the conjecture stated in the main text.

In particular, for the pair $(A_2,B_1)$, suppose that  $A_2$ measures observables corresponding to $W=\B_{W}\id+\str_{W}\bm \sigma \cdot \bm w$ and $W'=\B_{W'}\id+\str_{W'}\bm \sigma\cdot \bm w'$. Hence, the CHSH parameter for  $(A_2,B_1)$ follows from Eqs.~(\ref{chsh}) and~(\ref{measurement}) as
\begin{align}
	S(A_2,B_1)&=\an{W\otimes (Y+Y')}+\an{W'\otimes(Y-Y')} \nn \\
	&= \B_{W}\an{Y+Y'}+\str_W\an{\bm \sigma\cdot\bm w  \otimes (Y+Y')} \nn \\
	&~+\B_{W'}\an{Y-Y'}+\str_{W'}\an{\bm \sigma\cdot\bm w'  \otimes (Y-Y')}.
\end{align}
This is linear in the bias $\B_{W}$ and $\B_{W'}$, and hence, recalling that $\str+|\B|\leq1$, it achieves its extremal values for fixed strengths $\str_W,\str_{W}$ at $\B_{W}=\alpha(1-\str_W)$ and $\B_{W'}=\beta(1-\str_{W'})$, for $\alpha,\beta=\pm1$, yielding
\begin{align}
	S(A_2,B_1)\leq& \max_{\alpha,\beta} f_{\alpha\beta} \label{fab}
\end{align}
with
\begin{align}
	f_{\alpha\beta}&:=\alpha(1- \str_{W})\an{ Y+Y'}+\str_W\an{\bm \sigma\cdot\bm w  \otimes (Y+Y')} \nn \\
	&~+\beta(1-\str_{W'})\an{Y-Y'}+\str_{W'}\an{\bm \sigma\cdot\bm w' \otimes (Y-Y')}. \nn
\end{align}
Again, since $f_{\alpha\beta}$ is linear in the measurement strengths $\str_{W}, \str_{W'}$,  its extreme values must be achieved at $\str_{W}, \str_{W'} =0, 1$. Hence, we only need to analyse its values at these points. First, for $\str_{W}=\str_{W'}=0$ we have  $f_{\alpha\beta}=(\alpha+\beta)\an{Y}+(\alpha-\beta)\an{Y'}\leq |\alpha+\beta|+|\alpha-\beta|= 2$, and so the CHSH inequality cannot be violated for this choice. Second, for $\str_W=0$ and $\str_{W'}=1$ we have  \begin{align}
	f_{\alpha\beta}&=\alpha(\an{Y+Y'}+\an{\bm \sigma\cdot\bm w'  \otimes (Y-Y')}\nn\\
	& =2\alpha\big(\an{P_{\alpha\bm w'}\otimes Y}+\an{P_{-\alpha\bm w'}\otimes Y'}\big)\leq 2,\nn
\end{align} where $P_{\bm x}=\half(\id+\bm\sigma\cdot\bm x)$ denotes the projection onto unit spin direction $\bm x$, and so the CHSH inequality again cannot be violated for this choice, nor, by symmetry, for the choice $\str_W=1$ and $\str_{W'}=0$. Thus, it is only possible for $(A_2,B_1)$ to violate the inequality for the remaining choice $\str_{W}=\str_{W'}=1$, for which we have, via Eq.~(\ref{fab}),
\begin{align}
	S(A_2,B_1)&\leq\an{\bm w \cdot \bm \sigma \otimes (Y+Y')}+\an{\bm w' \cdot \bm \sigma \otimes (Y-Y')},
\end{align}
Note that equality holds for $W=\bm\sigma\cdot \bm w$ and $W'=\bm\sigma\cdot \bm w'$. Hence, $(A_2,B_1)$ can violate the CHSH inequality if and only if they can violate it via $A_2$  making projective measurements. A similar result holds for $(A_1,B_2)$ by symmetry.

Moreover, we can find the optimal projective measurements for $A_2$ to make (and similarly for $B_2$), as follows. First, for projective measurements $W=\bm\sigma\cdot \bm w$ and $W'=\bm\sigma\cdot \bm w'$, we have
\begin{align}
	S(A_2,B_1)&=\an{W \otimes (Y+Y')}+\an{W' \otimes (Y-Y')}\nn\\
	&= (\B_Y+\B_{Y'})\langle W\rangle + \an{W \otimes \bm \sigma \cdot(\tilde{\bm y}+\tilde{\bm y}') } \nn\\
	&~+(\B_Y-\B_{Y'})\langle W'\rangle + \an{W' \otimes \bm \sigma \cdot(\tilde{\bm y}-\tilde{\bm y}') } \nn\\
	&= \bm w\cdot \left[ (\B_Y+\B_{Y'})K\bm a + KT(\tilde{\bm y}+\tilde{\bm y}')\right] \nn\\
	&~+\bm w'\cdot \left[ (\B_Y-\B_{Y'})K\bm a + KT(\tilde{\bm y}-\tilde{\bm y}')\right] \nn\\
	&\leq \left|  (\B_Y+\B_{Y'})K\bm a + KT(\tilde{\bm y}+\tilde{\bm y}')\right| \nn\\
	&~+\left| (\B_Y-\B_{Y'})K\bm a + KT(\tilde{\bm y}-\tilde{\bm y}')\right|\nn\\
	&=S^*(A_2,B_1),
\end{align}
where $\tilde{\bm y}:=\str_Y \bm y$, $\tilde{\bm y}':=\str_{Y'} \bm y'$,  $\bm a$ is $A_1$'s Bloch vector for the initial shared state, and $T$ is the spin correlation matrix for the initial shared state. We have used the fact that $A_2$'s Bloch vector is $K\bm a$, and the spin correlation matrix for the state shared by $A_2$ and $B_1$ is $KT$ (see Sec.~II above). Equality holds in the last line by choosing $\bm w$ to be the unit vector in the $(\B_Y+\B_{Y'})K\bm a + KT(\tilde{\bm y}+\tilde{\bm y}')$ direction and $\bm w'$ to be the unit vector in the $(\B_Y-\B_{Y'})K\bm a + KT(\tilde{\bm y}-\tilde{\bm y}')$ direction.  Hence, $A_2$ and $B_1$ can violate the CHSH inequality if and only if $S^*(A_2,B_1)>2$, as claimed in the main text. It may similarly be shown that $A_1$ and $B_2$ can violate the CHSH inequality if and only if $S^*(A_1,B_2)>2$.

\subsection{B. Optimality of the singlet state for \\unbiased observables}
\label{sec:singlet}

Consider now the case that the observables are unbiased (i.e., $\B_X=\B_{X'}=\B_Y=\B_{Y'}=0$). Prior work~[3--22] has in fact been restricted to this case. We show that validity of the conjecture can then be reduced to testing it on the singlet state for a 9-parameter subset of observables.

First, using Eq.~(1) of the main text and Eqs.~(\ref{tx})--(\ref{txy}) above, it  follows for unbiased observables that the CHSH parameters for each pair $(A_j,B_k)$ are convex-linear in the spin correlation matrix $T$ of the initial state (and are independent of the Bloch vectors). 
Further, any physical spin correlation matrix $T$ can be written as a convex combination of the spin correlation matrices of maximally entangled states, as follows from the proof of Proposition~1 of~[36] (in particular, $T$ can be expressed as a mixture of the four Bell states corresponding to a basis in which $T$ is diagonal).

Now, any maximally entangled spin correlation matrix can be written as $T_{\rm me}=R'T_0R''^\top$, where $T_0=-I$ is the spin correlation matrix of the singlet state and $R', R''$  are local rotations of the first and second qubits. Hence, since the set of possible measurements is invariant under such rotations, it follows that searching the CHSH parameters over all measurements for a given $T_{\rm me}$ is equivalent to searching over all measurements for $T_0$, i.e, for the singlet state (corresponding to taking $\alpha=-\pi/4$ in Eq.~(12) of the main text). 

Moroever, noting that the singlet state is invariant under equal local rotations on each side, i.e., with $R'=R''$, the measurement direction $\bm x$ for $X$ can be fixed without loss of generality, as can the plane spanned by measurement directions $\bm x$ and $\bm x'$. Hence, the directions corresponding to $X,X',Y,Y'$ that need to be considered, for the purposes of the conjecture, form a 5-parameter set (the angle between $\bm x$ and $\bm x'$ in the given plane, and the angles specifying $\bm y$ and $\bm y'$).

Finally, for unbiased observables the only remaining free parameters are the four measurement strengths, $\str_{X}, \str_{X'}, \str_{Y}, \str_{Y'}$. Hence, the conjecture need only be tested for this case, whether numerically or analytically, for a 9-parameter subset of observables on a fixed maximally-entangled state, as claimed in the main text.

\section{IV. Numerical evidence for \\the conjecture}

As described in the main text, verification of our conjecture only requires consideration of the values of $S(A_1,B_1)$ and the proxy quantities $S^*(A_1,B_2)$, $S^*(A_2,B_1)$, and $S^*(A_2,B_2)$, in Eqs.~(10)--(12) of the main text, for the one-parameter set of pure initial states in Eq.~(8).  
These quantities are determined by a set of 17 parameters; 4 for each observable $X,X',Y,Y'$ as per Eq.~(3) of the main text, in addition to one for the state.

For the first case, illustrated in Fig.~2(a), the conjecture claims that the pairs $(A_1,B_1)$ and $(A_2,B_2)$ cannot both demonstrate CHSH Bell nonlocality. To test the conjecture for this case, we seek solutions to the problem
\begin{equation}
\begin{aligned}
 & \underset{\alpha,X,X',Y,Y'}{\text{max}} & & S^*(A_2,B_2)\\
& \hspace{14pt} \text{s. t.} & & |S(A_1,B_1)| \geq s,
\end{aligned}
\label{eq:passon_numerical_problem}
\end{equation}
where the quantities $S(A_1,B_1)$ and $S^*(A_2,B_2)$ are defined in Eqs.~(1) and (10) of the main text respectively, and the parameter $s\in[0,2\sqrt{2}]$ is fixed for each numerical test. 
For $s\geq2$, the conjecture requires that the solution to this problem does not exceed 2.
Varying $s$ allows investigation of the trade-off between the proxy quantities achievable by each pair of observers.
Finding a global optima for this problem is difficult, since it is does not have any particular structure which permits efficient solving in reasonable time.
Therefore, these numerical optimisations were performed using a constrained differential evolution (DE) solver~[33, 34] implemented in \textsc{SciPy}~[35]. The DE solver is a stochastic global search algorithm which operates by evolving a population of candidate solutions.
For our problem, each population member is a real-valued 17-dimensional vector, which is evolved by mutation, crossover, and selection processes, until a termination criteria is met~[33]. The constraints for the problem are handled using the approach detailed in~[34]. The population size, rate of mutation and crossover probability are control parameters chosen for each optimisation, which can impact convergence to a solution; see the package documentation~[35] for further details.

In Fig.~2(a), we sample 400 equally spaced values of $s$ from the interval $[0,2\sqrt{2}]$, and solve problem \eqref{eq:passon_numerical_problem} for each.
These are solved by the DE algorithm with a population size of $400$.  To help speed up convergence of the algorithm (particularly for large values of $s$), the initial population are chosen from of a Monte Carlo sample of 200 members satisfying the constraint in \eqref{eq:passon_numerical_problem} using an algorithm presented in~[36], in addition to 200 randomly generated vectors. The following solver parameters were chosen for this set of optimisations: $10^{-5}$ tolerance, `best1bin' strategy, 0.7 recombination rate and a dithering mutation rate sampled from [0.5,0.7] each iteration. Once the DE solver converged to a solution, a local least-squares optimizer was implemented to confirm the solution was a located at a local extremum. The solutions correspond to the blue data points in Fig.~2(a). It is evident that when $s\geq2$, the maximum value of the proxy quantity $S^*(A_2,B_2)$ never exceeds 2, supporting the monogamy conjecture for this grouping of observers.
For each fixed value of $s$, it should be noted that the numerical maximum is found when equality is attained by the constraint in \eqref{eq:passon_numerical_problem}.

Note that for $S(A_1,B_1)=s\leq2$, Fig.~2(a) indicates that it is always possible for the proxy quantity $S^*(A_2,B_2)$ to attain the maximum quantum value of $2\sqrt{2}$. 
This is indeed the case: suppose that $A_1$ and $B_1$ share a singlet state and measure the trivial observables $X=X'=Y=Y'=\B \id$, with bias $\B\in[-1,1]$ and strength $\str=0$, without disturbing the state. 
Then we have $S(A_1,B_1) = 2\B^2$, which ranges over all values of $[0,2]$.
Further, the state remains unchanged ($K=L=I$), so that $S^*(A_2,B_2) = 2\sqrt{2}$ can be achieved by performing the optimal CHSH measurements (this example is generalised in Sec.~V below). Conversely, however, Fig.~2(a) indicates that the range of $|S(A_1,B_1)|$ is restricted for values $S^*(A_2,A_B)\leq2$. This asymmetry is due to plotting the maximum values of the proxy quantity $S^*(A_2,B_2)$, which only depends on the choice of $X,X',Y,Y'$ (significantly reducing the number of search parameters), rather than the values of $S(A_2,B_2)$ (see also Sec.~V below).

Similar results were calculated for the pairs $(A_1,B_2)$ and $(A_2,B_1)$.
These are illustrated as blue points in Fig.~2(b).
Here, we solve the analogous problem
\begin{equation}
\begin{aligned}
 & \underset{\alpha,X,X',Y,Y'}{\text{max}} & & S^*(A_2,B_1) \\
& \hspace{14pt} \text{s. t.} & & S^*(A_1,B_2) \geq s, 
\end{aligned}
\label{eq:disordered_numerical_problem}
\end{equation}
where $S^*(A_1,B_2)$ and $S^*(A_2,B_1)$ are defined in Eqs.~(11) and (12) of the main text, and we confine $s$ to vary over $[2,2\sqrt{2}]$. 
The latter restriction is possible due to the symmetry of the problem under interchanging the roles of the Alices and  Bobs, which implies that the trade-off curve between the two proxy quantities must be symmetric about the line $S^*(A_1,B_2)=S^*(A_2,B_1)$.
Again, we solve this problem with the DE algorithm parameters listed above, this time with tolerance set to $10^{-8}$, for 500 equally spaced values of $s$.
The initial random population consisted of 400 members, one of which was chosen to correspond to the case where the pair $(A_1, B_2)$ achieve maximal CHSH violation on a singlet state, which significantly improved convergence time.
These results once again support the conjecture, since the extrema found for $S^*(A_2,B_1)$ never exceed $2$ over the range of $s$ (see Fig.~2(b)). 
Again, note that these numerical maximums are found when equality is satisfied for the constraint in \eqref{eq:disordered_numerical_problem}.

Finally, it is seen from Fig.~2 that the one-sided monogamy relations in Eqs.~(13) and~(14) of the main text do not hold for all possible choices of observables by $A_1$ and $B_1$, since some points lie above the red curves.  However, it is of interest to ask whether these relations might hold for the special case of unbiased observables, i.e., $\B_X=\B_{X'}=\B_Y=\B_Y=0$. 
For this case only the singlet state need be considered (see Sec.~III.B above), and  hence we randomly sampled  over  $10^{12}$ points for this state, to investigate this question. 
The results support a conjecture  that the one-sided monogamy relation $|S(A_1,B_1)|+S^*(A_2, B_2)\leq 4$ in Eq.~(14) holds for {\it all} unbiased observables, i.e., even without making  equal strength and/or orthogonality assumptions. 
In contrast, the monogamy relation $S^*(A_1, B_2)^2+S^*(A_2, B_1)^2\leq 8$  in Eq.~(13) was found to be numerically violated for some choices of unbiased observables that violate these assumptions

\section{V. One-sided monogamy relations} 
\label{sec:monog}

Eq.~(2) in the conjecture given in the main text is equivalent to the requirement that the CHSH parameters satisfy the general one-sided monogamy relations 
\begin{align}
	& \big| |S(A_1,B_1)|+|S(A_2,B_2)|-6\big|\nn\\
	&~~~+\big||S(A_1,B_1)|-|S(A_2,B_2)|\big|\geq 2, \\
	&\big||S(A_1,B_2)| + |S(A_2,B_1)| - 6 \big| +\nn\\
	&~~~+ \big| |S(A_1,B_2)| - |S(A_2,B_1)| \big|\geq 2. 
\end{align}
 In particular, the first relation rules out values of $S(A_1,B_1)$ and $S(A_2,B_2)$ that are both greater than 2, and the second relation similarly rules out values of $S(A_1,B_2)$ and $S(A_2,B_1)$ that are both greater than 2.  Note that these relations are saturated (up to the maximum values of $2\sqrt{2}$. For example, if $A_1$ and $B_1$ measure the trivial observables $X=X'=Y=Y'=\B\id$, then $S(A_1,B_1)=2\B^2$ ranges over $[0,2]$ while the lack of disturbance allows $|S(A_2,B_2)|$ to range over the full quantum range $[0,2\sqrt{2}]$. The converse result is obtained if instead $A_2$ and $B_2$ measure these trivial observables, and a similar saturation is obtained via $(A_1,B_2)$ and $(A_2,B_1)$ making trivial measurements.

It follows from the main text that the conjecture is also equivalent to the above relations with $S(A_1,B_2), S(A_2,B_1), S(A_2,B_2)$ replaced by the proxy quantities $S^*(A_1,B_2), S^*(A_2,B_1), S^*(A_2,B_2)$, 
corresponding to requiring the points in Fig.~2 of the main text to lie outside the shaded regions. 
 Here we derive the (less general but stronger) one-sided monogamy relations for the proxy quantities discussed in the main text. These hold for the cases of (i) unbiased observables, i.e., with
 \beq \label{zerobias}
 \B_X=\B_{X'}=B_Y=B_{Y'}=0, 
 \eeq
 (note that prior work~[3-22] is confined to this case), and/or (ii)  states with zero Bloch vectors, i.e., with
 \beq \label{bloch}
 \bm a=\bm b=\bm 0.
 \eeq 
(which includes all maximally entangled states), in combination with any of several mild measurement assumptions.

\subsection{A. One-sided monogamy relation for \\$(A_1,B_2)$ and $(A_2,B_1)$}

 Here we prove the monogamy relation in Eq.~(13) of the main text, i.e.,
\beq \label{monog1str}
S^*(A_1, B_2)^2 + S^*(A_2,B_1)^2 \leq 8 ,
\eeq
for each of the cases in Eqs.~(\ref{zerobias}) and~(\ref{bloch}), under the additional assumption of equal measurement strengths for each side.  We also prove this relation holds under the alternative additional assumption of orthogonal measurement directions for each side.   It follows that the proxy quantities $S^*(A_1, B_2)$ and $S^*(A_2, B_1)$ cannot both violate the CHSH inequality under  such restrictions. Hence, as per Sec.~III.A above, neither can both $S(A_1, B_2)$ and $S(A_2,B_1)$, thus confirming the conjecture under these restrictions.

\subsubsection{1. Convexity considerations}

To derive the monogamy relations, some convexity properties are needed to simplify the dependence of the quantities on the spin correlation matrices.

First, note  for either of the above cases in Eqs.~(\ref{zerobias}) and~(\ref{bloch}), that Eqs.~(11) and (12) of the main text simplify to
\begin{align} 
	S^*(A_1, B_2)&=|LT^\top(\tilde{\bm x}+\tilde{\bm x}')| + |LT^\top (\tilde{\bm x}-\tilde{\bm x}')|, \\
	S^*(A_2, B_1)&=|KT(\tilde{\bm y}+\tilde{\bm y}')| + |KT (\tilde{\bm y}-\tilde{\bm y}')|,
\end{align}
where $\tilde{\bm x}=\str_X\bm x$, etc. Importantly, these quantities are convex-linear with respect to the initial spin correlation matrix $T$. Further, the latter can always be written as a convex-linear combination $T=\sum_j w_jT_j$ of at most four spin correlation matrices $T_j$, corresponding to maximally-entangled states~[42] (specifically, to the four Bell states defined by the local basis sets in which $T$ is diagonal). Moreover, any maximally entangled state is related to the singlet state by local rotations, implying that $T_j=R_j' T_0 R_j''=-R'_j R_j''=:-R_j$, where $R_j', R_j''$ and $R_j=R_j' R_j''$ are rotation matrices, and $T_0=-I$ is the spin correlation matrix of the singlet state. 

Hence, since $|z|$ is a convex function,
\begin{align}
S^*(A_1, B_2)&\leq \sum_j w_j |LR_j^\top(\tilde{\bm x}+\tilde{\bm x}')| + |LR_j^\top (\tilde{\bm x}-\tilde{\bm x}')| \nn\\
&\leq \max_R \{ |LR^\top(\tilde{\bm x}+\tilde{\bm x}')| + |LR^\top (\tilde{\bm x}-\tilde{\bm x}')|\},
\label{simp12}
\end{align}	
where the maximum is over all rotations $R$, and similarly
\beq \label{simp21}
S^*(A_2, B_1)\leq  \max_R \{ |KR(\tilde{\bm y}+\tilde{\bm y}')| + |KR (\tilde{\bm y}-\tilde{\bm y}')| \} .
\eeq
These results will be used in obtaining Eq.~(\ref{monog1str}) for equal measurement strengths. 

Further, since $|z|^2$ is a convex function it also follows that
\begin{align}
&S^*(A_1, B_2)^2 + S^*(A_2,B_1)^2 \nn\\
&\leq \max_R \left\{ \left[ |KR(\tilde{\bm y}+\tilde{\bm y}')| + |KR (\tilde{\bm y}-\tilde{\bm y}')|\right]^2 \right.\nn\\
&\qquad~~~~ + \left. \left[ |LR^\top(\tilde{\bm x}+\tilde{\bm x}')| + |LR^\top (\tilde{\bm x}-\tilde{\bm x}')|  \right]^2 \right\} .
\label{simpsquare}
\end{align}
This result will be used in obtaining Eq.~(\ref{monog1str}) for orthogonal measurement directions.

\subsubsection{2. Equal strengths for each side}

Under the additional assumption that the measurements $X$ and $X'$ by $A_1$ have equal strengths, and similarly for the measurements $Y$ and $Y'$ by $B_1$, i.e.,
\beq 
\str_X=\str_{X'},~~ \str_Y=\str_{Y'}, \label{equalrev}
\eeq
Choosing $R$ to be the rotation saturating Eq.~(\ref{simp12}) then yields, 
\begin{align}
S^*(A_1, B_2)^2
	\leq & \str_X^2 \left(|LR^\top(\bm x+\bm x')|+ |LR^\top(\bm x-\bm x')|\right)^2\nn \\
	=& 4 \str_X^2  \left(|LR^\top \bm x_1|\cos\theta+ |LR^\top \bm x_2|\sin\theta \right)^2\nn \\
	\leq & 4 \str_X^2 \left(\bm x_1^\top R L^\top L R^\top \bm x_1+\bm x_2^\top RL^\top L R^\top \bm x_2\right) \nn \\
	\leq &  4 \str_X^2 \left[ s_1(L^\top L)+s_2(L^\top L)\right] \nn \\
	=&  \str_X^2 \left[(1+\R_Y)^2+(1+\R_{Y'})^2\right. \nn\\
	&~~~\left. +2(1-\R_Y)(1-\R_{Y'})(\bm y\cdot\bm y')^2 \right] \nn \\
	\leq &  \str_X^2 \left[(1+\R_Y)^2+(1+\R_{Y'})^2 \right. \nn\\
	&~~~\left. +2(1-\R_Y)(1-\R_{Y'}) \right] \nn \\
	\leq & \str_X^2 \left[(1+\R_Y)^2+(1+\R_{Y'})^2 \right. \nn\\
	&~~~\left. +(1-\R_Y)^2+(1-\R_{Y'})^2 \right] \nn \\
	=& 2\str_X^2 \left[2+ \R_Y^2+\R_{Y'}^2) \right].\label{Sa1b2}
\end{align}
Here, $2\cos\theta \,\bm x_1 := \bm x+\bm x'$ and $2\sin\theta\, \bm x_2:=\bm x -  \bm x'$ are orthogonal unit vectors defined via the half-angle $\theta$ between $\bm x$ and $\bm x'$ (implying that $R^\top\bm x_1$ and $R^\top \bm x_2$ are similarly orthogonal), and the singular values of $L^\top L$ (equivalent to the eigenvalues thereof) have been calculated via Eq.~(9) of the main text.
We similarly find, via Eq.~(\ref{simp21}), that
\beq
S^*(A_2, B_1)^2\leq 2\str_Y^2 \left[2+ \R_X^2+\R_{X'}^2 \right]. \label{Sa2b1}
\eeq

Finally, noting from the fundamental tradeoff relation~(\ref{rssum}) that 
\begin{align}
	\str_X\leq \min\{\rt{1-\R^2_X}, \rt{1-\R^2_{X'}}\},\\
	\str_Y\leq \min\{\rt{1-\R^2_Y}, \rt{1-\R^2_{Y'}}\},
\end{align}
Eqs.~(\ref{Sa1b2}) and~(\ref{Sa2b1}) yield
\begin{align}
	&S^*(A_1, B_2)^2+ S^*(A_2, B_1)^2 \nn \\
	&\leq  2\str_X^2 \left[(1+\R_Y^2)+(1+\R_{Y'}^2) \right]\nn \\
	&~~~+2 \str_Y^2\left[(1+\R^2_X)+(1+\R_{X'}^2) \right]. \nn  \\
	&\leq 2(1-\R^2_X)(1+\R^2_Y)+2(1-R_Y^2)(1+\R_{X}^2)\nn \\
	&~~~+2(1-\R^2_{X'})(1+\R^2_{Y'})+2(1-R_{Y'}^2)(1+\R_{X'}^2) \nn \\
	&=  4(1-\R^2_XR^2_Y)+4(1-\R^2_{X'}\R_{Y'}^2) \nn \\
	 &\leq 8,
\end{align}
as claimed in Eq.~(13) of main text and Eq.~(\ref{monog1str}) above.

\subsubsection{2. Orthogonal measurement directions for each side}

We now drop the equal strength assumption~(\ref{equalrev}), and instead assume that observables $X$ and $X'$ have orthogonal measurement directions, as do observables $Y$ and $Y'$, i.e.,  that
\beq \label{ortho}
\bm x \cdot \bm x'=0,\qquad \bm y \cdot \bm y'=0.
\eeq

We first show that we only need to consider the case where $\bm x,\bm x',R\bm y,R\bm y'$ lie in the same plane, for any rotation $R$ in Eq.~(\ref{simpsquare}). In particular, defining $\bm x'':=\bm x\times \bm x', \bm y'':=\bm y\times \bm y'$, note it follows from Eq.~(9) of the main text and the orthogonality condition~(\ref{ortho}) that
\beq
K =\frac{1+\R_{X'}}{2} \bm x\bm x^\top + \frac{1+\R_{X}}{2}\bm x'\bm x'^\top + \frac{\R_X+\R_{X'}}{2} \bm x''\bm x''^\top
\eeq
\beq \nn
L =\frac{1+\R_{Y'}}{2} \bm y\bm y^\top + \frac{1+\R_{Y}}{2}\bm y'\bm y'^\top + \frac{\R_Y+\R_{Y'}}{2} \bm y''\bm y''^\top .
\eeq
Hence, for a given rotation matrix $R$, one finds again using the orthogonality condition that
\begin{align}
|KR(\tilde{\bm y}\pm\tilde{\bm y}')|^2 &= \frac14\left\{ (1+\R_{X'})^2[\bm x\cdot  R(\tilde{\bm y}\pm\tilde{\bm y}')]^2 \right. \nn\\
&~~~+ (1+\R_{X})^2[\bm x'\cdot  R(\tilde{\bm y}\pm\tilde{\bm y}')]^2 \nn\\
&~~~+ \left. (\R_X+\R_{X'})^2[\bm x''\cdot  R(\tilde{\bm y}\pm\tilde{\bm y}')]^2 \right\} . 
\label{kr}
\end{align}
Since $\R_X,\R_{X'}\leq 1$ it follows that the third term has the smallest weighting factor, so that $|KR(\tilde{\bm y}\pm\tilde{\bm y}')|$ is maximised for any $R$ by choosing directions such that $\bm x''$ is orthogonal to $R(\tilde{\bm y}\pm\tilde{\bm y}')$, i.e., such that $\bm x,\bm x'$ lie in the same plane as $R\bm y, R\bm y'$. One similarly finds that $|LR^\top(\tilde{\bm x}\pm\tilde{\bm x}')|$ is maximised by choosing directions such that $\bm y,\bm y'$ lie in the same plane as $R^\top\bm x, R^\top \bm x'$, i.e., again such that $\bm x,\bm x'$ lie in the same plane as $R\bm y, R\bm y'$.  Note that the latter two vectors are also orthogonal to each other.

Thus, choosing $R$ to be the rotation saturating Eq.~(\ref{simpsquare}), and introducing the parameter $\beta$ to characterise the relative angles between (coplanar) $\bm x, \bm x'$ and $R\bm y, R\bm y'$, i.e., 
\beq
R\bm y =\bm x\cos \beta  + \bm x'\sin \beta,~R\bm y' =\bm x\sin \beta - \bm x'\cos \beta .
\eeq
and
\beq
\bm x = R\bm  y\cos \beta  + R\bm y'\sin \beta,~\bm x' = R\bm y\sin \beta - R\bm y'\cos \beta ,
\eeq
we find via Eq.~(\ref{kr}) that 
\begin{align}
	|KR(\tilde{\bm y}\pm\tilde{ \bm y}')|^2&= \frac14 (1+\R_{X'})^2[\bm x\cdot  R(\str_Y\bm y\pm\str_{Y'}\bm y')]^2 \nn\\
	&+ \frac14 (1+\R_{X})^2[\bm x'\cdot  R(\str_Y\bm y\pm\str_{Y'}\bm y')]^2\nn \\
	&=\frac14 (1+\R_{X'})^2 (\str_Y\cos\beta\pm\str_{Y'}\sin\beta)^2 \nn\\ &+\frac14 (1+\R_{X})^2 (\str_Y\sin\beta\mp\str_{Y'}\cos\beta)^2.
\end{align}
Hence, using ($(a+b)^2 \leq (a+b)^2+(a-b)^2=2(a^2+b^2)$ and the fundamental tradeoff relation~(\ref{rssum}) yields
\begin{align}
	&(|KR(\tilde{\bm y}+\tilde{ \bm y}')|+|KR(\tilde{\bm y}-\tilde{ \bm y}')|) ^2 \nn\\
	&\leq 2 (|KR(\tilde{\bm y}+\tilde{ \bm y}')|^2+| KR(\tilde{\bm y}-\tilde{ \bm y}')|^2) \nn \\
	&= (1+\R_{X'})^2 (\str_Y^2\cos^2\beta+\str_{Y'}^2\sin^2\beta) \nn\\
	&~~+ (1+\R_{X})^2 (\str_Y^2\sin^2\beta + \str_{Y'}^2\cos^2\beta) \nn \\
	&\leq  (1+\R_{X'})^2 (1-\R_Y^2) \cos^2\beta + (1+\R_{X'})^2 (1-\R_{Y'}^2)\sin^2\beta \nn \\
	&~~+ (1+\R_{X})^2 (1-\R_Y^2)\sin^2\beta + (1+\R_{X})^2 (1-\R_{Y'}^2) \cos^2\beta\nn \\
	&= (1+\R_X)^2  (1-\R_Y^2) + (1+\R_{X'})^2 (1-\R_{Y'}^2) \nn \\
	&~~+\left[(1+\R_{X'})^2-(1+\R_{X})^2\right] (\R_{Y'}^2-\R_{Y}^2) \cos^2\beta . \label{SSa2b1}
\end{align}
Similarly, we obtain
\begin{align}
	&(|LR^\top(\tilde{\bm x}+\tilde{\bm x}')| + |LR^\top (\tilde{\bm x}-\tilde{\bm x}')|)^2 \nn \\
	&\leq (1-\R_{X}^2)(1+\R_Y)^2+(1-\R_{X'}^2)(1+\R_{Y'})^2 \nn \\
	&~~+(\R_{X'}^2-\R_X^2)\left[(1+\R_{Y'})^2-(1+\R_{Y})^2\right] \cos^2\beta. \label{SSa1b2}
\end{align}
Substituting Eqs.~(\ref{SSa2b1}) and (\ref{SSa1b2}) into Eq.~(\ref{simpsquare}) then gives
\begin{align}
	&S^*(A_1, B_2)^2 + S^*(A_2, B_1)^2 \nn \\
	&\leq  (1-\R_Y^2)(1+\R_X)^2+(1-\R_{Y'}^2)(1+\R_{X'})^2 \nn \\
	&~~+(1-\R_{X}^2)(1+\R_Y)^2+(1-\R_{X'}^2)(1+\R_{Y'})^2 \nn \\
	&~~+(\R_{Y'}^2-\R_{Y}^2)\left[(1+\R_{X'})^2-(1+\R_{X})^2\right] \cos^2\beta \nn \\
	&~~+(\R_{X'}^2-\R_X^2)\left[(1+\R_{Y'})^2-(1+\R_{Y})^2\right] \cos^2\beta \nn\\
	&= P + (\R_X-\R_{X'})(\R_{Y}-\R_{Y'})Q\cos^2\beta,
	\label{pq}
\end{align}
where
\begin{align}
	P:= & (1-\R_Y^2)(1+\R_X)^2+(1-\R_{Y'}^2)(1+\R_{X'})^2 \nn \\
	&+(1-\R_{X}^2)(1+\R_Y)^2+(1-\R_{X'}^2)(1+\R_{Y})^2 
\end{align}
and 
\begin{align}
	Q:= &(\R_X+\R_{X'}+2)(\R_{Y}+\R_{Y'})\nn\\
	& + (\R_X+\R_{X'})(\R_{Y}+\R_{Y'}+2).
\end{align}

Finally, noting that $P,Q\geq0$, if 
$(\R_X-\R_{X'})(\R_{Y}-\R_{Y'}) \leq 0$,
then Eq.~(\ref{pq}) is maximised for $\cos^2\beta=0$, implying
\begin{align}
	&S^*(A_1, B_2)^2 + S^*(A_2, B_1)^2 \nn \\
	&\leq  (1-\R_Y^2)(1+\R_X)^2+(1-\R_{Y'}^2)(1+\R_{X'})^2 \nn \\
	&~~+(1-\R_{X}^2)(1+\R_Y)^2+(1-\R_{X'}^2)(1+\R_{Y'})^2 \nn \\
	&\leq  2\max_{\R_X,\R_Y}\left[(1-\R_Y^2)(1+\R_X)^2+(1-\R_{X}^2)(1+\R_Y)^2 \right] \nn \\
	& = 8, \label{ABA'B'}  
\end{align}
while if $(\R_X-\R_{X'})(\R_{Y}-\R_{Y'}) \geq 0$,
then Eq.~(\ref{pq}) is maximised for $\cos^2\beta=1$, implying
\begin{align}
	&S^*(A_1, B_2)^2 + S^*(A_2, B_1)^2 \nn \\
	&\leq  (1-\R_{Y'}^2)(1+\R_X)^2+(1-\R_{X}^2)(1+\R_{Y'})^2 \nn \\
	&~~+(1-\R_{Y'}^2)(1+\R_X)^2+(1-\R_{X}^2)(1+\R_{Y'})^2 \nn \\
   	&\leq 2\max_{\R_X,\R_{Y'}}\left[(1-\R_{Y'}^2)(1+\R_X)^2+(1-\R_{X}^2)(1+\R_{Y'})^2\right] \nn \\
	&= 8. \label{AB'A'B}  
\end{align}
Thus, in either case we again obtain the one-sided monogamy relation in  Eq.~(\ref{monog1str}),  as claimed in the main text.

\subsection{B. One-sided monogamy relation for \\$(A_1,B_1)$ and $(A_2,B_2)$}

We now prove the one-sided monogamy relation in Eq.~(14) of the main text, under the combined assumptions of unbiased observables, and equal strengths and orthogonal measurement directions for each side. In fact, for this combination the upper bound can be improved to
\beq
|S(A_1,B_1)|+S^*(A_2,B_2) \leq \frac{16}{3\sqrt{2}}\sim 3.7<4,
\eeq
as noted in the main text. Hence, $S(A_1, B_1)$ and the proxy quantity $S^*(A_2, B_2)$ cannot both violate the CHSH inequality, implying that neither can both $S(A_1, B_1)$ and $S(A_2,B_2)$ (see main text), thus confirming the conjecture for this case. One-sided monogamy relations for  more general cases, in which either of the equal strength or orthogonality assumptions is dropped, will be derived in a forthcoming paper~[26], via a significant generalisation of the Horodecki criterion. 

First, under the above assumptions it follows from Sec.~III.B that the result needs only to be proved for the singlet state, i.e, for $T=-I$ and $\bm a=\bm b=0$.  But for this state the CHSH parameter for the pair $(A_1,B_1)$ can be calculated  via Eqs.~(1) and~(3) of the main text and Eq.~(\ref{zerobias}), (\ref{equalrev}) and~(\ref{ortho}) above to give
\begin{align}
|S(A_1,B_1)|&=\str_X \str_Y| \bm x\cdot\bm y +  \bm x\cdot\bm y' + \bm x'\cdot\bm y  -  \bm x'\cdot\bm y'| \nn\\
&\leq \str_X\str_Y \left[ |\bm x\cdot (\bm y +  \bm y')|+ |\bm x'\cdot (\bm y -  \bm y')|\right] \nn\\
&\leq\str_X\str_Y \left[ |\bm y +  \bm y'|+ |\bm y -  \bm y'|\right] \nn\\
&=2\sqrt{2}\str_X\str_Y \nn\\
&=  2\sqrt{2}\,\sqrt{1-\R_X^2}\,\sqrt{1-\R_Y^2} , \label{s11bound}
\end{align}
where the last line follows by noting that the fundamental tradeoff relation~(\ref{rssum}) is saturated for unbiased observables. 

Moreover, from the Horodecki criterion in Eq.~(9) of the main text, it follows for the singlet state that
\begin{align}
S^*(A_2,B_2)^2&\leq  4[s_1(KL)^2 + s_2(KL)^2]\nn\\
&\leq 4[s_1(K)^2s_1(L)^2 +s_2(K)^2s_2(L)^2],
\end{align}
using Theorem~IV.2.5 of Ref.~[43]. Further, the matrices $K$ and $L$ follow from Eq.~(9) of the main text under the equal strengths assumption as (again noting the saturation of tradeoff relation~(\ref{rssum}))
\beq
K=\R_XI_3 + \half(1-\R_X)(\bm x\bm x^\top + \bm x'\bm x'^\top),
\eeq
\beq
L=\R_YI_3 + \half(1-\R_Y)(\bm y\bm y^\top + \bm y'\bm y'^\top).
\eeq
The orthogonality assumption then yields $s_1(K)=s_2(K)=\half(1+\R_X)$ and $s_1(L)=s_2(L)=\half(1+\R_Y)$ by inspection, and so
\begin{align}
	S^*(A_2,B_2)^2&\leq \half (1+\R_X)^2 (1+\R_Y)^2. \label{s22bound}
\end{align}

Finally, defining $f(x)=2\sqrt{1-x^2}$ and $g(x):=1+x$, Eqs.~(\ref{s11bound}) and~(\ref{s22bound}) give
\begin{align}
	&|S(A_1,B_1)| + |S^*(A_2,B_2)|\nn\\
	&\qquad\leq  \frac{ f(\R_X)f(\R_Y)+g(\R_X)g(\R_Y)}{\sqrt{2}} \nn\\
	&\qquad \leq\frac{\sqrt{f(\R_X)^2+g(\R_X)^2}\sqrt{f(\R_Y)^2+g(\R_Y)^2} }{\sqrt{2}} \nn\\
	&\qquad \leq \max_\R \frac{f(\R)^2+g(\R)^2}{\sqrt{2}} \nn\\
	&\qquad= \max_\R \frac{16- (1-3\R)^2}{3\sqrt{2}} \nn\\
	&\qquad= \frac{16}{3\sqrt{2}},
\end{align}
as claimed, where the second inequality follows via $\bm m\cdot\bm n\leq |\bm m||\bm n|$ for $\bm m=(f(\R_X),g(\R_X))$ and $\bm n=(f(\R_Y),g(\R_Y))$. Note that the maximum is achieved for
\beq
\str_X=\str_Y =\frac{2\sqrt{2}}{3}\sim 0.943, ~~~ \R_X=\R_Y=\frac13,
\eeq
and the optimal CHSH directions.

\section{VI. Biased measurement selections}

Recall that, in the example of the main text,  $X$ and $Y$ are each selected with probability $1-\epsilon$ and measured with strengths $\str_X=\str_Y=\str$ and reversibilities $\R_X=\R_Y=\R$;   $X'$ and $Y'$ are selected with probability $\epsilon$ and are projective, with strengths $\str_{X'}=\str_{Y'}=1$ and reversibilities $\R_{X'}=\R_{Y'}=0$; and the measurement directions correspond to the optimal CHSH directions~[27], i.e., $\bm x$ and $\bm x'$ are orthogonal with $\bm y=(\bm x+\bm x')/\sqrt{2}$ and $\bm y'=(\bm x-\bm x')/\sqrt{2}$. 

The biasing of the measurement selections modifies the average matrices $K$ and $L$ in Eq.~(\ref{barka}) to
\begin{align} 
K_\epsilon &:= (1-\epsilon)K^X+\epsilon K^{X'}\nn\\ 
&= (1-\epsilon)[ \R I  + (1-\R) \bm x\bm x^\top ]   
+ \epsilon  \bm x'\bm x'^\top ,
\label{keps} \\
L_\epsilon  &:= (1-\epsilon)L^Y+\epsilon L^{Y'}\nn\\ 
&= (1-\epsilon)[ \R I  + (1-\R) \bm y\bm y^\top ]   
+ \epsilon  \bm y'\bm y'^\top , \label{leps}
\end{align}
as per the main text.

For the optimal CHSH directions this gives, in the $\{\bm x,\bm x',\bm x\times\bm x'\}$ basis,
\beq
K_\epsilon= \begin{pmatrix}
	1-\epsilon & 0 &0\\ 0 & \R+\epsilon(1-\R)  &0 \\ 0 & 0 & (1-\epsilon)\R
\end{pmatrix} ,
\eeq
\beq
L_\epsilon= \begin{pmatrix}
	\frac{1+(1-\epsilon)\R}{2} & \frac{1-(1-\epsilon)\R}{2}-\epsilon & 0 \\\frac{1-(1-\epsilon)\R}{2}-\epsilon & \frac{1+(1-\epsilon)\R}{2} & 0\\ 0 & 0 & (1-\epsilon)\R
\end{pmatrix}  .
\eeq
Only the principal $2\times2$ submatrices of $K_\epsilon$ and $L_\epsilon$ contribute to the two largest singular values of $K_\epsilon L_\epsilon$ (the smallest value is $(1-\epsilon)\R$), implying $s_1(K_\epsilon L_\epsilon)^2+s_2(K_\epsilon L_\epsilon)^2$ is given by the trace of the $2\times2$ principal submatrix of  $(K_\epsilon L_\epsilon)(K_\epsilon L_\epsilon)^\top=K_\epsilon L_\epsilon^2 K_\epsilon$, which may be evaluated to give 
\begin{align}
s_1(K_\epsilon L_\epsilon)^2+s_2(K_\epsilon L_\epsilon)^2 &= \half \left[1+\R^2 -2\epsilon(1-\R+\R^2)\right.\nn\\
&\qquad~ \left. +\epsilon^2(2-2\R+\R^2) \right]^2 .
\end{align}
It immediately follows via Eq.~(10) of the main text that $A_2$ and $B_2$ can violate the CHSH inequality only if
\begin{align}
S^*(A_2,B_2) &= \sqrt{2} \left[1+\R^2 -2\epsilon(1-\R+\R^2)\right.\nn\\
&\qquad~~~ \left. +\epsilon^2(2-2\R+\R^2) \right] >2 ,
\end{align}
which is equivalent to
\beq \label{rminus}
\R >\R_-(\epsilon) := \frac{[\sqrt{2}-(1-\epsilon)^2]^{1/2}-\epsilon}{1-\epsilon} ,
\eeq
as per Eq.~(16) of the main text. This is monotonic increasing in $\epsilon$, implying in particular that one must have
\beq 
\R>\R_-(0)=(\sqrt{2}-1)^{1/2} \sim 0.64359 .
\eeq
This corresponds, via tradeoff relation~(\ref{rssum}), to the upper bound
\beq \label{smax}
\str<\str_{\max}:= \sqrt{1-\R_-(0)^2} =(2-\sqrt{2})^{1/2}\sim 0.7654.
\eeq
on the strength $\str$, as noted in the main text.

Note that the requirement $\R_-(\epsilon)\leq1$ implies that the probability $\epsilon$ of selecting a projective measurement $X'$ or $Y'$ cannot be arbitrarily large. In particular, Eq.~(\ref{rminus}) can be inverted to give
\beq \label{epsr}
\epsilon_-(\R)=\frac{1-\R+ \R^2-\sqrt{\sqrt{2}(2-2 \R+  \R^2)-1}}{2 -2  \R+ \R^2} ,
\eeq
which is a monotonic increasing function of $\R$, and hence violation of the CHSH inequality by $A_2$ and $B_2$ is only possible at all if
\beq
\epsilon< \epsilon_-(1)=1-(\sqrt{2}-1)^{1/2} \sim 0.356406 .
\eeq
Moreover, for {\it both} pairs $(A_1,B_1)$ and $(A_2,B_2)$ to violate the CHSH inequality one further requires that the strength satisfy
\beq \label{smin}
\str> \str_{\min}:=8^{1/4}-1\sim 0.6818,
\eeq 
as per Eq.~(16) of the main text, which together with tradeoff relation~(\ref{rssum}) and Eq.~(\ref{rminus}) implies that $\R$ and $\epsilon$ must satisfy
\beq \label{range}
\R_-(\epsilon) <\R < \R_+ ,
\eeq
with 
\beq
\R_+:=\sqrt{1-\str_{\min}^2} =  2^{3/4}\sqrt{2^{1/4}-1} \sim 0.7315.
\eeq

Eq.~(\ref{range}) can clearly be satisfied for sufficiently small values of $\epsilon$, i.e., for sufficiently large selection biases (since $\R_-(0)<\R_+$). Hence, it is possible to have two-qubit recycling of Bell nonlocality for sufficiently biased measurement selections. The maximum possible value of $\epsilon$ that allows such a violation of one-sided monogamy follows from Eqs.~(\ref{epsr}) and~(\ref{range}) as
\beq
\epsilon_{\max}=\epsilon_-(\R_+) \sim 0.0794626\sim 7.9\% ,
\eeq
as per Eq.~(17) of the main text.

Finally, as noted in the main text, the pairs $(A_1,B_2)$ and $(A_2,B_1)$ can also violate a CHSH inequality under the above conditions. This  expected, on the grounds that these share qubits of which only one qubit has been recycled following measurement, so that it should be even easier for them to violate a Bell inequality than  it is for $(A_2,B_2)$, as both qubits of the latter pair have been recycled. 
We demonstrate this explicitly below by considering the optimal case that $\epsilon$ approaches zero.

In particular, in this limit $K$ and $L$ in Eqs.~(10) and~(11) of the main text are replaced by $K_0$ and $L_0$ in Eqs.~(\ref{keps}) and~(\ref{leps}). Further, for the optimal CHSH directions one finds
\beq
K_0\tilde{\bm y}=\frac{S}{\sqrt{2}}\begin{pmatrix} 1\\ \R \end{pmatrix},~~
K_0\tilde{\bm y}'=\frac{1}{\sqrt{2}}\begin{pmatrix} 1\\ -\R \end{pmatrix} ,
\eeq
and hence, noting that the Bloch vectors $\bm a$ and $\bm b$ vanish for the singlet state, Eq.~(10) of the main text is replaced by
\begin{align}
S^*(A_1,B_2) =&\frac{1}{\sqrt{2}}\left| \begin{pmatrix} 1+\str \\ -\R(1-\str) \end{pmatrix} \right| + \frac{1}{\sqrt{2}}\left| \begin{pmatrix} 1-\str \\ -\R(1+\str) \end{pmatrix} \right|\nn \\
=&  \frac{1}{\sqrt{2}} \left[ (1+\str)^2+\R^2(1-\str)^2\right]^{1/2} \nn\\
&+\frac{1}{\sqrt{2}} \left[ (1-\str)^2+\R^2(1+\str)^2\right]^{1/2}  
\end{align}
in the limit $\epsilon\rightarrow0$. Using the tradeoff relation~(\ref{rssum}) then gives
\begin{align}
S^*(A_1,B_2)\leq & \frac{1}{\sqrt{2}} \left[ (1+\str)^2+(1-\str^2)(1-\str)^2\right]^{1/2} \nn\\
&+ \frac{1}{\sqrt{2}} \left[ (1-\str)^2+(1-\str^2)(1+\str)^2\right]^{1/2}
\end{align}
in this limit. 
One similarly finds
\beq
L_0\tilde{\bm x}=\frac{S}{2}\begin{pmatrix} 1+\R \\ 1-\R \end{pmatrix},~~
L_0\tilde{\bm x}'=\frac{1}{2}\begin{pmatrix} 1-\R\\ 1+\R \end{pmatrix},
\eeq
which yields the same upper bound for $S^*(A_2,B_1)$ via via Eq.~(11) of the main text. 

Hence, both pairs can violate a CHSH inequality if this upper bound is greater than 2, corresponding to
\beq \str < \str_0 := \sqrt{\sqrt{3}-1} \sim 0.8556,
\eeq
or equivalently to
\beq
\R > \R_0:= \sqrt{2-\sqrt{3}} \sim 0.5176.
\eeq
Noting that $\str_0>\str_{\max}$ in Eq.~(\ref{smax}, it follows that both pairs $(A_1,B_2)$ and $(A_2,B_1)$ can violate the CHSH inequality  if $(A_2,B_2)$ can, as claimed.

\end{document}